%% file: main.tex
\documentclass[twoside]{IEEEtran}

\usepackage{amsfonts,amsmath,amssymb,mathtools,amsthm}
\usepackage{color}
\usepackage{cite}
\usepackage{hyperref}
\hypersetup{backref, colorlinks=true, linkcolor=blue, citecolor=blue, urlcolor = blue}
\usepackage{ushort}
\usepackage[a4paper, total={7in, 10in}]{geometry}
\input{macro.tex}

\makeatletter
\newcommand\footnoteref[1]{\protected@xdef\@thefnmark{\ref{#1}}\@footnotemark}
\makeatother

\title{\Large \bf
Rate-cost tradeoffs in continuous-time control\\ with a biomolecular application 
}

\author{Yorie Nakahira, Fangzhou Xiao, Victoria Kostina, John C. Doyle \thanks{Y.~Nakahira (\href{mailto:ynakahir@andrew.cmu.edu}{ynakahir@andrew.cmu.edu}) is with Carnegie Mellon University, USA.  F.~Xiao (\href{mailto:xiaofangzhou@westlake.edu.cn}{xiaofangzhou@westlake.edu.cn}) is with  Westlake University, China. V.~Kostina and J. C. Doyle (\href{mailto:vkostina@caltech.edu}{vkostina@caltech.edu}, \href{mailto:doyle@caltech.edu}{doyle@caltech.edu}) are with California Institute of Technology, USA.
    This work was supported in part by the National Science Foundation (NSF)
under grant CCF-1751356. A preliminary version of this work was presented at ACC 2018~\cite{nakahira2018biomolecular}, in a paper that did not contain proofs.
    } }

\begin{document}

\maketitle
\thispagestyle{empty}
\pagestyle{empty}

\begin{abstract}

This paper focuses on rate-limited control of the generalized Ornstein-Uhlenbeck process where the control action can be either multiplicative or additive, and the noise variance can depend on the control action. We derive a lower bound on the data rate necessary to achieve the desired control cost. The lower bound is attained with equality if the control is performed via an additive white Gaussian channel. The system model approximates the dynamics of a discrete-state molecular birth-death process, and the result has direct implications on the control of a biomolecular system via chemical reactions, where the multiplicative control corresponds to the degradation rate, the additive control corresponds to the production rate, and the control objective is to decrease the fluctuations of the controlled molecular species around their desired concentration levels.

\end{abstract}


\vspace{-3mm}
\section{Introduction}
Since the rise of systems biology and synthetic biology at the beginning of this century, a central objective has been to understand how biomolecular systems in a cell perform complex control objectives, such as cell differentiation, homeostasis, and accurate timing of cell cycle~\cite{alonBook}. Recent years have seen fruitful applications of established frameworks from control theory to biological systems~\cite{doyle2011glyco,doyle2014heartrate,murrayBook,briat2016antithetic}. The connection between biological systems and engineered ones became even more apparent with the advent of large-scale engineered hardware and software systems~\cite{doyle2002,doyle2011neuro}.

A major challenge in studying biomolecular systems is the difficulty of constructing complete system models~\cite{hilfinger2016identify,paulsson2016oscillation}. A cell usually contains a network of species, and the species of interest usually depends on the states of other species, which in turn depends on the states of the remaining species in the network. To deal with this type of modeling complexities, Lestas et al.~\cite{lestas2010fundamental} pioneered an information-theoretic approach to biomolecular control. They considered the problem of driving the number of molecules of a chemical species \(X\) towards the desired level while suppressing the fluctuations in that number. The control is effected via the production rate of \(X\). By treating other parts of the network, which may contain other molecular species whose production rate depends on \(X\),  as a communication channel, and using causal rate-distortion function~\cite{gorbunov1974prognostic}, Lestas et al.~\cite{lestas2010fundamental} derived a lower bound on the variance of the number of molecules of \(X\) in terms of the capacity of that channel.

This work extends the approach of~\cite{lestas2010fundamental}
to account for control via the degradation rate, since \cite{lestas2010fundamental} only considers control via the production rate.  
Our contribution falls under the realm of control under
communication constraints. The necessary and sufficient data rate through the feedback loop in order to achieve system stability in linear stochastic control is studied in~\cite{yuksel2010stochastic,nair2004stabilizability,tatikonda2004control}. The optimal controller structure, separation principles, performance bounds are studied in~\cite{bar1974dual,TatikondaSahaiMitter,charalambous2008lqg,freudenberg2010stabilization,silva2011framework,yuksel2014jointly,tanaka2017lqg,silva2016characterization,kostina2016rate}. Notable results include separation principle between the controller design and communication protocols~\cite{fischer1982optimal,fu2012lack,TatikondaSahaiMitter}, the link between anytime capacity and stabilizability~\cite{sahai2006necessity}, and the relation between optimal cost and the causal rate-distortion function~\cite{TatikondaSahaiMitter,silva2011framework,kostina2016rate,gorbunov1973nonanticipatory,bucy1980distortion,derpich2012improved,rezaei2006rate,charalambous2014nonanticipative}. 
The concept of directed mutual information was introduced by Massey~\cite{massey1990causality}, who showed that the capacity of feedback channels is upper-bounded in terms of a maximal directed information problem. 
We model the number of molecules as a generalized Ornstein-Uhlenbeck process where the control can be additive (control via the production rate) or multiplicative (control via the degradation rate).
This is a continuous-time process, while most prior works in control under communication constraints discretize the time. Causal rate-distortion function for both discrete- and continuous-time Gauss-Markov processes was derived in~\cite{gorbunov1974prognostic}. Continuous-time directed mutual information was introduced in~\cite{weissman2013directed}.

The rest of the paper is organized as follows. Section~\ref{sec:setup} sets up the nonlinear continuous-time control problem over a communication channel, introduces the informational rate-cost function and the channel capacity in that context, and shows that in order for the desired control cost to be achievable, it is necessary that the rate-cost function evaluated at that cost does not exceed the channel capacity, extending a known converse result to continuous time.  Section~\ref{sec:main} presents our main result - a lower bound on the rate-cost function of a generalized Ornstein-Uhlenbeck process with additive and multiplicative control actions. Section~\ref{sec:biomolecular} discusses how our result applies to biomolecular control.

\subsubsection*{Notation} $\mathbb Z_+$ denotes the set of non-negative integers, and $\reals_+$ denotes the set of non-negative reals. $\mathcal N(\mu, \sigma^2)$ denotes a Gaussian random variable with mean $\mu$ and variance $\sigma^2$. 
All logarithms are assumed to be natural. For a continuous time process $\left\{ X(t) \right\}_{t \in \reals_+}$, we use the notations $X^u \triangleq \left\{ X(t) \right\}_{t \in [0, u]}$, $X^{u-} \triangleq \left\{ X(t) \right\}_{t \in [0, u)}$, and $X_{t}^{u} \triangleq \left\{ X(s) \right\}_{s \in [t, u]}$; \emph{stationary mean} and \emph{variance} are denoted by 
\begin{align}\label{eq:smean}
\mean[X] \triangleq \lim_{t \rightarrow \infty} \E[X(t)], \;\;
\svar[X] \triangleq \lim_{t \rightarrow \infty} \var[X(t)]
\end{align}
provided that the said limits exist.   
For a discrete time process $\left\{ X[k] \right\}_{k \in \ints_+}$, we use the notations $X^n \triangleq \left\{ X[k] \right\}_{k \in \left\{0, 1, \ldots, n\right\}}$ and $X^{k}_{n} \triangleq \left\{ X[i] \right\}_{i \in \left\{k, k + 1, \ldots, n\right\}}$. The stationary mean and variance of discrete-time processes are defined analogously to \eqref{eq:smean}. Round brackets are used to represent continuous-time stamps (e.g., $\X(k\delta)$), while square brackets are used to represent discrete-time indices (e.g., $\X[k]$). 

\vspace{-3mm}
\section{Continuous-time control}
\label{sec:setup}
\subsection{Operational problem setup}
For a pair of discrete-time sequences $\left\{X[k],Y[k]\right\}_{k = 1}^n$, a causally conditional kernel (or probability distribution) is defined as~\cite{kramer2003capacity} $\p_{Y^n \| X^{n}} \triangleq \prod_{k = 1}^n \p_{Y[k] |  Y^{k-1} , X^{k}}$, $\p_{Y^n \| X^{n-1}} \triangleq \prod_{k = 1}^n \p_{Y[k] |  Y^{k-1} , X^{k-1}}$ Equivalently, one can identify $\p_{Y^n \| X^{n}}$ with the collection of conditional probabilities $\left\{\p_{Y[k] |  Y^{k-1} , X^{k}}\right\}_{ k = 1}^n$, and  $\p_{Y^n \| X^{n-1}}$ with  $\left\{\p_{Y[k] |  Y^{k-1} , X^{k-1}}\right\}_{ k = 1}^n$. Likewise, for a pair of continuous-time processes $\left\{X(t), Y(t) \right\}_{t =0}^T$, we identify causally conditional kernels $\p_{Y^T \| X^{T}}$ and $\p_{Y^T \| X^{T-}}$ with the collections of transition probabilities $\left\{\p_{Y(t) | \X^t, Y^{t-}}\right\}_{ t= 0}^T$ and $\left\{\p_{Y(t) | \X^{t-}, Y^{t-}}\right\}_{ t= 0}^T$, respectively. 

\begin{figure}[h]
\center
\includegraphics[width=0.3\textwidth]{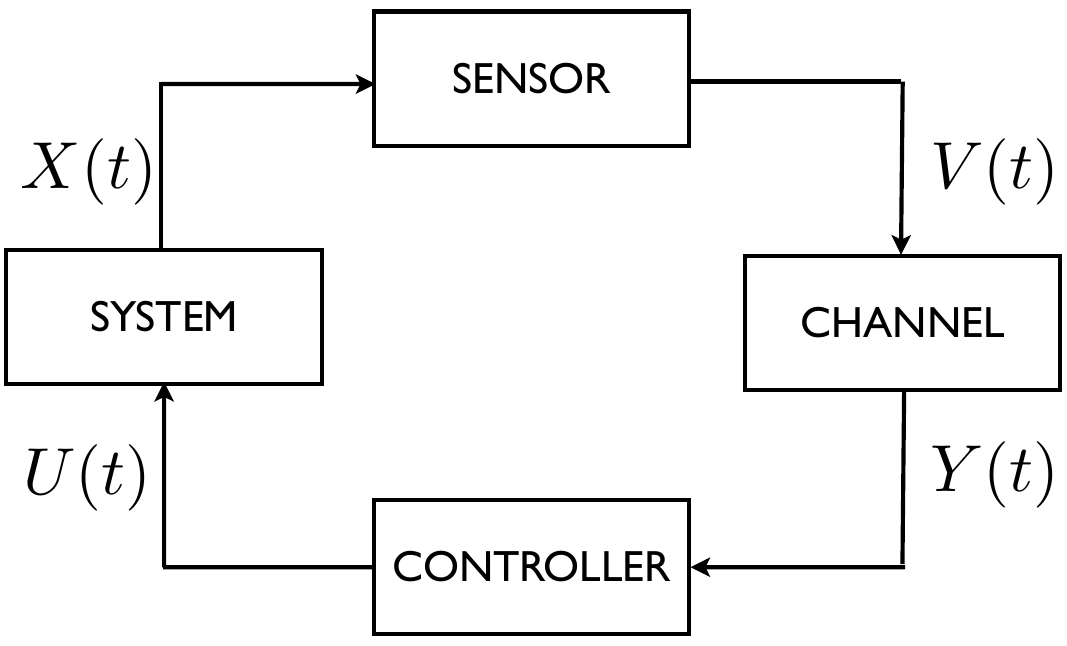}
\caption{The continuous-time control system.}
\label{fig:system}
\end{figure}

In Fig.~\ref{fig:system}, the stochastic continuous-time \emph{dynamical system} is the collection $\left\{ \p_{X^t \| U^{t-}} \right\}_{t \in \reals_+}$, where $X(t)$ is the system state and $U(t)$ is the control signal, and 
the stochastic continuous-time \emph{communication channel} is the kernel $\left\{P_{Y^t \| V^{t}}\right\}_{t \in \reals_+}$, where $V(t)$ is the channel input and $Y(t)$ is its output. 
The joint distribution of the stochastic processes in Fig.~\ref{fig:system} is given by
\begin{align}
\p_{X^t , V^t, Y^t, U^t} &= \p_{X^t \| U^{t-}} \p_{V^t \| X^{t}, Y^{t-}} \p_{Y^t \| V^{t}} \p_{U^t \| Y^{t}} .
\label{eq:systemjoint}
\end{align}

The sensor $\left\{\p_{V^t \| X^{t}, Y^{t-}}\right\}$ and the controller $\left\{P_{U^t \| Y^{t}}\right\}_{t \in \reals_+}$ seek to center the stationary mean of $\left\{X(t)\right\}_{t \in \mathbb R_+}$ around a desired value $\xgoal$ subject to a constraint on the stationary variance, \ie, (recall \eqref{eq:smean}) 
\begin{align}
\label{eq:main_const}
&\mean[ \xdev ] = \xgoal, & \svar[  \xdev ]  \leq D.
\end{align}
Based on the parameters of the dynamical system and the communication channel, the control objective \eqref{eq:main_const} may or may not be attainable. The amount of randomness in the dynamical system is quantified by its \emph{rate-cost function}, while the quality of the channel - by its \emph{capacity}. We proceed to define these quantities and to show that a necessary condition for \eqref{eq:main_const} to be attainable is the channel capacity being no smaller than the system's rate-cost function (Proposition~\ref{prop:total}, below).

\subsection{Directed mutual information}
\label{sec:dirmi}
For a pair of discrete-time sequences $\left\{X[k],Y[k]\right\}_{k = 1}^n$, the discrete-time directed information is defined as~\cite{massey1990causality} $I(X^n \rightarrow Y^n) \triangleq \sum_{k = 1}^{n} I ( X^k ; Y[k] | Y^{k-1})$, where the conditional mutual information is $I(X;Y|Z) \triangleq \mathbb E_{(X, Y, Z) \sim P_{Y|XZ} P_{XZ}} \left[ \log \frac{ d P_{Y|XZ}(Y|XZ)} {d P_{Y|Z}(Y|Z) } \right]$.
The continuous-time directed information is defined as follows.

\begin{defn}[{Directed information in continuous time~\cite[eq. (25)]{weissman2013directed}}]
\label{def:directedinfo}
Let $\left\{X(t), Y(t) \right\}_{t =0}^\infty$ be a pair of stochastic processes.
The \emph{directed information} from $\left\{X(t)\right\}_{t = 0}^T$ to $\left\{Y(t)\right\}_{t = 0}^T$ is defined as
\begin{align}
\label{eq:def_directedinfo}
&I ( X^T \rightarrow Y^T ) \triangleq \lim_{ \de \rightarrow 0 } \sum_{k = 1}^{ \lfloor T/\de \rfloor} I (  X^{k \de} ; Y_{(k-1)\de}^{k \de}  | Y_{0}^{(k -1)\de}).
 \end{align} 
 The \emph{directed information rate} is the quantity $ \lim_{T \to \infty} \frac 1 T I ( X^T \rightarrow Y^T )$. 
 \end{defn}
 The next definition extends the causally conditioned directed information $I(X^n \rightarrow  Y^n \| Z^{n-1}) \triangleq \sum_{k = 1}^{n} I ( X^k ; Y[k] | Y^{k-1}, Z^{k-1})$~\cite{kramer2003capacity} to continuous time. 
 \begin{defn}[Causally conditioned directed information in continuous time]
 \label{def:causaldirectedinfo}
 Let $\left\{X(t), \hat X(t), Y(t)\right\}_{t=0}^\infty$ be stochastic processes.
 The directed information from $\left\{X(t)\right\}_{t = 0}^T$ to $\left\{Y(t)\right\}_{t = 0}^T$ causally conditioned on $\left\{Z(t)\right\}_{t = 0}^{T-}$ is defined as
\begin{align}
 &I (   X^T \rightarrow Y^T   \| Z^{T-} ) \notag\\
 \triangleq& \lim_{ \de \rightarrow 0 }  \sum_{k = 1}^{ \lfloor T/\de \rfloor } I ( X^{k \de}; Y_{(k-1)\de}^{k \de}  \| Y_0^{(k-1) \de } , Z_0^{(k-1) \de}  ) .
 \label{eq:def_directedinfocond}
 \end{align}
  The \emph{causally conditioned directed information rate} is the quantity $ \lim_{T \to \infty} \frac 1 T (   X^T \rightarrow Y^T   \| Z^{T-} ) $. 
\end{defn}

\subsection{Rate-cost function}

\begin{defn}[Continuous-time rate-cost function] 
\label{def:rate_cost_control_continuous}
The rate-cost function of the continuous-time system $\left\{\p_{X^t \| U^{t-}}\right\}_{t \in \reals_+}$ is defined as
\begin{equation}
\begin{aligned}
\mathbb R(D) \triangleq &  \inf \limsup_{T \rightarrow \infty}  \frac{ 1 } {T } I ( \xdev^T \rightarrow U^T  ),
\end{aligned}
\end{equation}
where the infimum is over all\footnote{In the sequel, we will impose regularity constraints on the set of allowable policies, see \eqref{eq:ubound}, \eqref{eq:4bound}. This will not affect the applicability of the general results in Section~\ref{def:directedinfo}. \label{fn:reg}}
 control policies $\left\{\p_{U^t \|\xdev^t}\right\}_{t \in \reals_+}$  achieving control objective \eqref{eq:main_const}. 
\end{defn}
The rate-cost function characterizes the minimum amount of information flow through the feedback loop required to attain control objective \eqref{eq:main_const}. 
The discrete-time rate-cost function is a proxy for the bit rate required to encode the control actions to achieve control objective~\cite[Sec. II.C]{kostina2016rate}, and serves as a lower bound to the minimum required channel capacity~\cite[Prop.~1]{kostina2016rate}.   

Definition~\ref{def:rate_cost_control_continuous} extends discrete-time rate-cost function~\cite[Def. 1]{kostina2016rate} to continuous time. This paper simplifies the control cost to the mean-square deviation from the target, while~\cite[Def. 1]{kostina2016rate} considers the more general linear quadratic regulator cost. 
Definition~\ref{def:rate_cost_control_continuous} tightens\footnote{Convergence of a sequence implies convergence of its Ces\`aro means.} the constraint on the Ces\`aro-mean variance in~\cite[Def. 1]{kostina2016rate}  to the constraint \eqref{eq:main_const} on the stationary variance of $\left\{X(t)\right\}_{t \in \reals_+}$; to accommodate the stationary variance constraint, it also places the infimum over control policies in front of the $\limsup$. The stationary variance constraint is easier to deal with in the context of time-varying systems considered in this paper.

The next definition concerns the causal compression of controlled dynamical systems and quantifies the bitrate needed to achieve distortion $D$ in the representation of the system state.
\begin{defn}[Continuous-time rate-distortion function] 
\label{def:rate_cost_estimation_continuous}
Fix a continuous-time system $\left\{\p_{X^t \| U^{t-}}\right\}_{t \in \reals_+}$.
The rate-distortion function is defined as
\footnote{The subscript  `e' in $\mathbb R_e(D)$ stands for  `estimation'.}
\begin{equation}
\begin{aligned}
\mathbb R_e(D) \triangleq  & \inf \limsup_{T \rightarrow \infty}  \frac{ 1 } {T }  I (   \xdev^T \rightarrow \hX^T   \| U^{T-} ), 
\end{aligned}
\end{equation}
where the infimum is over all$^{\ref{fn:reg}}$ estimation $\left\{\p_{\hX^t \| X^t, U^{t-}}\right\}_{t \in \reals_+}$ and control $\left\{\p_{U^t \| \hat X^t }\right\}_{t \in \reals_+}$ policies achieving 
\begin{align}
 \E [  ( \X - \hX  ) ^2 ]   \leq D.
\label{eq:staterr}
\end{align}
\end{defn}
The rate-distortion function is the minimum directed information rate from the system state to the state estimate required to sustain the estimation accuracy.

The conditioning on $U^{t-}$ in the optimization variable $\p_{\hX^t \| X^t, U^{t-}}$ signifies that the past control actions are known to the estimator, which is the case if the estimator is co-located with the controller. Definition~\ref{def:rate_cost_estimation_continuous} extends~\cite[Def. 4]{kostina2016rate} to continuous time. 
 Similar to~\cite[Def. 4]{kostina2016rate}, the past controls in Definition~\ref{def:rate_cost_estimation_continuous} are considered available side information in the compression process, but dissimilar to~\cite[Def. 4]{kostina2016rate}, Definition~\ref{def:rate_cost_estimation_continuous} also optimizes over all possible past control sequences $\left\{\p_{U^t \| \hat X^t }\right\}_{t \in \reals_+}$ in order to determine the best tradeoffs achievable for the purpose of estimation, not control. This extra optimization is immaterial for linear systems, where the additive control signal has no effect on estimation~\cite[Prop.~2]{kostina2016rate}. Theorem~\ref{thm:Rd}, below, reveals other mechanisms of control that do not affect the rate-distortion function, e.g., multiplicative control policies with fixed stationary mean.

\begin{prop}
\label{prop:continuous-time-separation}
The rate-cost function cannot be smaller than the rate-distortion function: 
\begin{align}
\label{eq:br}
\mathbb R(D)  \geq \mathbb R_e(D).
\end{align}
\end{prop}
The proof of Proposition \ref{prop:continuous-time-separation} is in Appendix \ref{app:prop1}.
Proposition \ref{prop:continuous-time-separation} continues to hold for discrete-time systems.  For example, for the scalar Gaussian discrete-time system $X[k+1] = a X[k] + U[k] + V[k]$, we have $\mathbb R_e(D) = \mathbb R(a^2 D + \sigma_V^2) \leq \mathbb R(D)$, where $\sigma_V^2$ is the variance of $V[k]$~\cite[Cor.~1]{kostina2016rate}, with strict inequality in the nontrivial regime $(1 - a^2) D < \sigma_V^2$. In contrast, for the continuous-time system, the estimator's output is incorporated into the control action immediately, rather than at the next discrete time step, and the lower bound \eqref{eq:br} can be attained with arbitrary precision. 

\subsection{Channel capacity}

\begin{defn}[Continuous-time channel capacity~\cite{weissman2013directed}]
The feedback capacity of the communication channel $\left\{P_{Y^t \| V^{t}}\right\}_{t \in \reals_+}$ is defined as 
\begin{align}
\label{eq:channel_capacity}
\capa \triangleq \sup \liminf_{T \rightarrow \infty} \frac{1}{T} I( V^T \rightarrow Y^T ) , 
\end{align}
where the supremum is taken over all$^{\ref{fn:reg}}$ feedback communication policies $\left\{P_{V^t \| Y^{t-}}\right\}_{t \in \reals_+}$.
\label{defn:capacity}
\end{defn}
The feedback capacity is the maximum directed information rate from the channel input to the channel output. 

The following impossibility result extends~\cite[Proposition~1]{kostina2016rate} to nonlinear systems and to continuous time. 
\begin{prop}\label{prop:total}
Consider the control system \eqref{eq:systemjoint} with plant $\left\{\p_{X^t \| U^{t-}}\right\}_{t \in \reals_+}$ controlled over a channel $\left\{\p_{Y^t \| V^{t}}\right\}_{t \in \reals_+}$. In order to achieve mean-square deviation $D$ \eqref{eq:main_const} from the target, it is necessary that
\begin{align}
\label{eq:R-and-Y}
&\mathbb R(D)  \leq \capa.
\end{align}
\end{prop}

The proof of Proposition \ref{prop:total} is in Appendix~\ref{app:prop1}.
Usually, equality is not attained in \eqref{eq:R-and-Y} because of the conflicting objectives of minimizing the directed information to attain the rate-cost function and maximizing it to attain the channel capacity. Furthermore, it has been argued~\cite{sahai2006necessity} that channel capacity is too lenient a characteristic of the channel for control purposes because the exponential growth of an uncontrolled unstable system places a stringent constraint on the exponential decay of the transmission error probability. Nevertheless, \eqref{eq:R-and-Y} provides a converse bound on the quality of the channel necessary to attain the control objective.   

\section{Main result}
\label{sec:main}
We consider the model of the dynamics of $\left\{X(t)\right\}_{t \in \mathbb R_+}$ described by the stochastic differential equation (SDE)
 \begin{align}
 \label{eq:ou}
 d X =    (\birthx- \deathx X)dt +\sigma dW,
\end{align}
 where $\birthx(t) \geq 0$, $\deathx(t) \geq 0$, $\sigma(t) \geq 0$, and $\left\{W(t) \right\}_{t \in \mathbb R_+}$ is a Wiener process. 
 Control enters the system in \eqref{eq:ou} via the additive action $\f$, the multiplicative action $\A$, and the magnitude of the noise $\sigma$, i.e.
 \begin{align}
U(t) \triangleq \left\{\A(t),  \f(t), \sigma(t) \right\}.
\label{eq:U}
\end{align}
The dynamical system in \eqref{eq:ou} extends the standard linear Gaussian model (the Ornstein-Uhlenbeck process) to incorporate the multiplicative control action via $\deathx(t)$ and the dependence of the noise variance $\sigma^2(t)$ on the control actions. 

We limit our attention to the class of control policies such that multiplicative control $\left\{\A(t)\right\}_{t \in \reals_+}$ is uniformly bounded and $\left\{\sigma(t)\right\}_{t \in \reals_+}$ has bounded 4th stationary moment, 
\begin{align}
 |\A(t)| &\leq \A_{\max},  \label{eq:ubound} \\
\E[ \sigma^4] &< \infty  \label{eq:4bound} 
\end{align}
for some constant $\A_{\max}$, and the stationary mean $\E[\A]$ exists.

 \begin{thm}
The rate-distortion function of the system in \eqref{eq:ou} is lower-bounded as
\begin{align}
\label{eq:br-form}
\mathbb R_e(D) \geq  \frac{ \E \left[ \sigma^2 \right] }{2 D} - \E[\mu].
\end{align}
The lower bound is achieved if and only if $\frac 1 D \geq \mathrm{ess} \limsup_{t \to \infty} \frac{1 - \mu(t)^2}{\sigma(t)^2}$. Furthermore, the policy that achieves it is a linear Gaussian policy.  
\label{thm:Rd}
\end{thm}

\begin{proof}
Sampling the continuous-time system \eqref{eq:ou} at uniformly spaced intervals, we obtain a discretized system that is Gaussian conditioned on the past control actions. This observation allows us to compute its rate-distortion in closed form (Theorem~\ref{thm:rddiscrete} in Appendix~\ref{app:thm1} below). Towards that end, we first derive a lower bound on the distortion-rate function of a single transmission of a random variable that is  Gaussian conditioned on side information available at both encoder and decoder (Theorem~\ref{thm:grd} in Appendix~\ref{app:thm1} below). We then develop a recursion linking the distortion at step $k$ to that at step $k - 1$, and we apply the single-transmission bound at each step. The bound is achieved by the Kalman filter applied to the output of an auxiliary additive noise Gaussian channel with the noise power set to achieve the stationary mean-square error $D$.  
See Appendix \ref{app:thm1} for details.
\end{proof}

Due to Proposition \ref{thm:Rd}, the rate-cost function is also bounded by the right-hand side of \eqref{eq:br-form}. 
Particularizing Theorem~\ref{thm:Rd} to the Gauss-Markov process (constant $\mu$ and $\sigma$ in \eqref{eq:ou}) recovers the result of  Gorbunov and Pinsker~\cite[(1.61)]{gorbunov1974prognostic} employed in~\cite[Eq. (39)]{lestas2010fundamental} to model molecular control via the production rate $\lambda$. 
Theorem~\ref{thm:Rd} extends this classical result to the scenario where 
$\left\{\deathx(t)\right\}_{t \in \mathbb R_+}$ and $\left\{\sigma^2(t)\right\}_{t \in \mathbb R_+}$ are stochastic processes. 

Using Proposition \ref{thm:Rd} and Proposition~\ref{prop:total}, we rewrite \eqref{eq:br-form} in terms of the minimum channel capacity $\mathbb C$ required to sustain stationary variance $D = \Var{X}$ as 
\begin{align}
\Var{X} \geq \frac{\E \left[ \sigma^2 \right]}{ 2 \left( \mathbb C + \E\left[ \mu \right] \right) }. 
\label{eq:rdcbound}
\end{align}
 
\section{Biomolecular control}
\label{sec:biomolecular}
The evolution of the number of molecules $\left\{X(t)\right\}_{t \in \mathbb R_+}$ is modeled as a birth-death process, defined as follows~\cite[Ch. 17]{feller1950introduction}.

\begin{itemize}
\item $\left\{X(t)\right\}_{t \in \reals_+}$ is a continuous-time Markov chain with state space $\mathbb Z_+$, i.e.,  $X(t) \in \mathbb Z_+$.
\item The transition probability from state $n \in \mathbb Z_+$ to state $n + m \in \mathbb Z_+$ satisfies, as $h \to 0$,
\begin{align}
\nonumber
&\Pro \left[X(t+\dt) = n + m | X(t) = n \right] \\
\label{eq:transition_prob}
&= \begin{dcases}
\birthx(t) \dt + o(\dt) & \text{ if } m = 1\\
\deathx(t)n \dt + o(\dt) & \text{ if }  m = -1\\
o(\dt) & \text{ if }  |m| > 1,
\end{dcases}
\end{align}
\ck{where $\birthx(t) \geq 0$ is the \emph{production rate}, and $\deathx(t) \geq 0$ is the \emph{degradation rate}}. 
\end{itemize}

The reaction rate of the degradation process is governed by the multiplication of the degradation rate $\deathx$ and the number of molecules, according to the mass action law~\cite{voit2015plos}.

If the number of molecules $X(t)$ is large, the number of birth events of species $X$ in the time interval $[t , t + \dt]$ is approximately $\mathcal N(\birthx (t) \dt,\, \birthx(t) \dt)$; likewise, the number of death events is approximately $\mathcal N(\deathx(t) X(t) \dt,\, \deathx(t) X(t) \dt)$~\cite{gillespie2000langevin}.   Under this approximation, the discrete-state birth-death process $\left\{X(t)\right\}_{t \in \mathbb R_+}$ simplifies to the continuous-state SDE \eqref{eq:ou} with $\sigma^2(t) =   \birthx(t)  + \deathx(t) X(t)$, also known as the \textit{chemical Langevin equation}~\cite{gillespie2000langevin}. Further approximating the $X(t)$ in the noise variance $\sigma^2(t)$ by $\E\left[X\right]$, we set in \eqref{eq:ou} \footnote{Lestas et al. \cite{lestas2010fundamental} approximates $\sigma^2(t)$ by the constant $\E\left[\sigma^2\right]$; we here refine the approximation to \eqref{eq:sigma_w}.}
 \begin{align}\label{eq:sigma_w}
 \sigma^2(t) =   \birthx(t)  + \deathx(t) \frac{\E \left[ X \right]}{\gamma_X}, 
 \end{align}
 where $\gamma_X$, the \textit{degradation efficiency} of \(X\), is defined as
\begin{align}
\label{eq:one_def}
&\one \triangleq \frac{ \mean[X]\mean[\deathx] } {\mean[ \deathx X ]}. 
\end{align}
 If the degradation rate process $\left\{\mu(t)\right\}_{t \in \mathbb R_+}$ is independent of the process $\left\{X(t)\right\}_{t \in \mathbb R_+}$, then $\one = 1$. Otherwise, the degradation efficiency $0 < \gamma_X < \infty$ is determined by the statistical dependence between those processes. 

The controller (usually a combination of other molecular processes in the cell) acts on the molecular process $\left\{X(t)\right\}_{t \in \mathbb R_+}$ either through the production rate $\birthx(t)$ or through the degradation rate $\deathx(t)$. Thus, $\left\{\birthx(t), \deathx(t)\right\}_{t \in \mathbb R_+}$ is a stochastic process, and $\left\{X(t)\right\}_{t \in \reals_+}$ is a causal random transformation of that stochastic process. Thus, molecular control via the production and/or the degradation rate falls under the purview of Theorem~\ref{thm:Rd} (under the continuous SDE approximation described in the previous paragraph).

To summarize, the evolution of the number of molecules is modeled as \eqref{eq:ou}. Control policies satisfying \eqref{eq:sigma_w} (as well as \eqref{eq:ubound}, \eqref{eq:4bound} as before) for a fixed $\ell_X$ are allowed. Control objective is to minimize 
\begin{align}
F_X \triangleq \frac{\Var{X}}{\mean[X] },
\label{eq:fano}
\end{align}
known as the \emph{stationary Fano factor} of the random process $\left\{X(t)\right\}_{t \in \mathbb R_+}$. It is a normalized measure of distributional spread and is equal to 1 for the Poisson distribution.

The converse result \eqref{eq:rdcbound} is rewritten in terms of biologically relevant quantities as follows (Appendix~\ref{apx:biomolecular}): 
\begin{align}
F_X \geq \frac{1}{\ell_X \mathbb C + \gamma_X}.
 \label{eq:rdcbio}
\end{align}
Here, $\ell_X$ is the average lifetime of the $X$ molecules, $\ell_X$, related to the stationary mean via Little's queuing formula~\cite{little1961proof}
\begin{align}
\mean[X] = \ell_X \mean[ \lambda].
\label{eq:little}
\end{align}

The bound in \eqref{eq:rdcbio} extends~\cite[Eq. (14)]{lestas2010fundamental} to the scenario where control is performed via the degradation rate, i.e., $\gamma_X \neq 1$. 
The form of \eqref{eq:rdcbio} begs the question of whether, for the same channel capacity $\mathbb C$, control via the degradation rate can achieve a lower Fano factor than  control via the production rate alone. Indeed, although the degradation efficiency \eqref{eq:one_def} can be increased by making $\mu$ and $X$ negatively correlated since $\E\left[\mu X\right] = \mathrm{Cov}\left(\mu, X\right) + \E\left[\mu\right] \E\left[X\right]$, the channel limits the controller's ability to do this, since the controller does not have a direct view of $X(t)$.  At an extreme, if the channel capacity is 0, the controller does not have the ability to match $\mu$ to $X$ because it does not observe $X$, and only $\gamma_X = 1$ is achievable. Furthermore, if the channel is an additive white Gaussian noise (AWGN) channel, i.e., $dY = Vdt + dB$, where $\left\{B(t) \right\}_{t \in \mathbb R_+}$ is a Wiener process, then \eqref{eq:rdcbio} is attained with equality, and there is nothing to gain by controlling  the degradation rate (Appendix~\ref{apx:biomolecular}):
\vspace{-1mm}
\begin{prop}
If the channel is an AWGN channel, then
 \begin{align}
F_X = \frac{1}{\ell_X \mathbb C + 1}
\label{eq:fanogauss}
\end{align}
is the minimum achievable Fano factor.
\label{prop:Fano}
\end{prop}

\section{Conclusion}
This paper sets up a general problem of continuous-time control over a communication channel and shows a general converse result (Proposition~\ref{prop:total}) on the communication requirements necessary to attain control objective in terms of the rate-cost function (Definition~\ref{def:rate_cost_control_continuous}) of the process describing the system dynamics. It introduces an extension of the Ornstein-Uhlenbeck process with both additive and multiplicative control actions and derives a lower bound to its rate-cost function (Theorem~\ref{thm:Rd}). Theorem~\ref{thm:Rd} leads to a new converse bound \eqref{eq:rdcbio} on the stationary Fano factor of the birth-death process describing the number of molecules where the control is performed via either the production or the degradation rate. Optimal control over AWGN channels is achieved via the production rate, in which case the converse is attained with equality (Proposition~\ref{prop:Fano}). For control over more general noisy channels, it remains an open question of whether increasing degradation efficiency through negative correlation between degradation rate and the number of molecules of the controlled species is a viable strategy.

\appendices

\addcontentsline{toc}{chapter}{APPENDICES}

\section{Proof of Proposition \ref{prop:continuous-time-separation} and Proposition \ref{prop:total}}
\label{app:prop1}

\begin{proof}[Proof of Proposition \ref{prop:continuous-time-separation}] For each $P_{U^T\| \hat X^T}$, consider the estimation policy  
\begin{align}
\label{eq:prop1_defyhat}
\hX(t) \triangleq \E [  \X(t)  | \hX^{t-} , U^{t-}   ].
\end{align} 
Due to Jensen's inequality, 
\begin{align}
\label{eq:var_bound2}
 \E [ ( \X (t)    - \hX(t) )^2   ] \leq \var [ \X(t) ].
\end{align}
Now, the result follows from
\begin{align}
I (   \X^T \rightarrow \hX^T   \| U^{T-} ) \leq I ( \xdev^T \rightarrow U^T  ),
\label{eq:mi_bound2}
\end{align}
which is due to 
$ I ( \xdev^T \rightarrow U^T  ) = I ( \xdev^T \rightarrow (U^T, \hat \xdev^T  )) \geq  I (   \X^T \rightarrow \hX^T   \| U^{T-} )$, where the lower bound is due to the directed mutual information chain rule~\cite[(3.14)--(3.16)]{kramer1998PhD} and the equality holds because $\hat X(t)$ \eqref{eq:prop1_defyhat} is a deterministic function of  $\hX^{t-} , U^{t-}$.
\end{proof}

\begin{proof}[Proof of Proposition \ref{prop:total}] According to  data processing for directed information across the feedback loop~\cite[Lemma 1]{tanaka2017lqg}, 
\begin{align}
 I(X^T \to U^T) \leq I(X^T \to Y^T \| U^{T-}).
\end{align}
This result, shown in~\cite[Lemma 1]{tanaka2017lqg} for discrete time, naturally generalizes to continuous time by taking limits in the definition of continuous-time directed information \eqref{eq:def_directedinfo}, \eqref{eq:def_directedinfocond}. 
Since $P_{X^{T} Y^T \| U^{T-} V^T} = P_{X^T \| U^{T-}} P_{Y^T \| V^T}$, we have by the (conditional) directed mutual information chain rule~\cite[(3.14)--(3.16)]{kramer1998PhD} $I((X^T, V^T) \to Y^T \| U^{T-}) = I(V^T \to Y^T \| U^{T-}) + I(X^T \to Y^T \| U^{T-}, V^T) = I(V^T \to Y^T \| U^{T-}) $, which leads to
\begin{align}
 I(X^T \to Y^T \| U^{T-}) \leq  I(V^T \to Y^T \| U^{T-}).
\end{align}
Finally, we have $  I(V^T \to Y^T \| U^{T-}) \leq \max_{u} I(V^T \to Y^T \| U^{T-} = u)$, and the result follows after maximization over $\left\{P_{V^T \| Y^{T-}, U^{T-} = u}\right\}$ using  $P_{Y^T V^T \| U^{T-}} = P_{Y^T \| V^T} P_{V^T \| Y^T, U^{T-}}$. 
\end{proof}

\section{Proof of Theorem \ref{thm:Rd}}
\label{app:thm1}
We convert the continuous-time system into a discrete-time system and then construct an optimal estimator for the discrete-time system. This estimator converges to the optimal estimator for the continuous-time system as the sampling interval approaches zero.

\subsection{Rate-distortion function for the discrete-time system}

Fix time horizon $T > 0$ and a sampling interval $\delta > 0$. Sampling system \eqref{eq:ou} at times $\left\{k \delta\right\}_{k = 1}^{\lfloor  T/ \de \rfloor}$, we obtain the following discrete-time system:
\begin{align}
\label{eq:discrete-dynamics}
\X[k+1] =  \A [k] \X[k] + \f[k] + W[k],
\end{align} where $\X[k] \triangleq \X(k \delta)$ is the value of $\X(t)$ at the sampled time $t = k \delta$, $W[k] \sim \mathcal N( 0,\sigma[k]^2)$ are independent, and
\begin{align}
\label{eq:A_sigma:a}
&\A [k] \triangleq \exp \left\{ -\int_{(k - 1) \delta}^{k \delta}  \A(\tau) d\tau \right\},\\
\label{eq:A_sigma:b}
&\f[k] \triangleq  \int_{(k - 1) \delta}^{k \delta} \exp  \left\{ -\int_{\tau}^{k \delta} \A(v ) dv \right\} \f(\tau) d\tau,\\
\label{eq:A_sigma:Y}
&W[k] \triangleq  \int_{(k - 1) \delta}^{k \delta} \exp  \left\{ -\int_{\tau}^{k \delta} \A(v ) dv \right\} W(\tau) d\tau,\\
\label{eq:A_sigma:d}
&\sigma[k]^2 \triangleq \int_{(k - 1) \delta}^{k \delta} \exp \left\{ -2 \int_{\tau}^{k \delta} \A(v ) dv \right\} \sigma(\tau)^2 d\tau  .
\end{align}

As its continuous-time counterpart \eqref{eq:ou}, the system in \eqref{eq:discrete-dynamics} may be controlled via $\A [k]$, $\f[k]$, and $\sigma[k]$, i.e., 
\begin{align}
U[k] \triangleq \left\{\A[k],  \f[k], \sigma[k] \right\}. 
\label{eq:Udiscrete}
\end{align}
The control objective is \eqref{eq:main_const}: to steer the system towards a target while keeping the stationary variance of $\left\{X[k]\right\}$ at $D$.

While the system \eqref{eq:discrete-dynamics} is more general than the standard Gaussian linear system, where $U[k] =  \f[k]$ and the noise variance $\sigma[k]^2$ is not affected by the control, the conditional distribution  $\p_{X^k \| U^{k-1}}$ is still Gaussian. This will allow us to compute its rate-distortion function in closed form. 

The rate-distortion function for a discrete-time system  $\left\{\p_{X^k \| U^{k-1}}\right\}_{k =1}^\infty$ is defined as 
\begin{align}
 \label{eq:cost-rate-formula:d} 
 \mathbb R_{e,\delta}(D) &\triangleq   \inf \limsup_{\T \rightarrow \infty} \frac{ 1 } {\T }  I ( \X^\T \rightarrow \hX^\T \| U^{\T-1} ) ,
\end{align}
where the infimum is over all$^{\ref{fn:reg}}$ estimation $\left\{\p_{\hX^k \| X^k, U^{k-1}}\right\}_{k =1}^\infty$ and control $\left\{\p_{U^k \| \hat X^k }\right\}_{k =1}^\infty$ policies satisfying \eqref{eq:main_const} (cf. Definition~\ref{def:rate_cost_estimation_continuous}; see also~\cite[Def. 4]{kostina2016rate}).

\begin{thm}
The rate-distortion function \eqref{eq:cost-rate-formula:d} for the system in \eqref{eq:discrete-dynamics} with control in \eqref{eq:Udiscrete} and distortion constraint in \eqref{eq:main_const} is lower-bounded as
 \begin{align}
\!\!\!\! \mathbb R_{e, \delta}(D)  \geq \limsup_{n \to \infty} \frac{1}{2 n} 
 \sum_{k = 1}^{n} \E\left[  \log \left(  \A[k]^2 + \frac {\sigma[k]^2}{D}\right) \right],
 \label{eq:Rdlb}
\end{align}
The lower bound is achieved if and only if $\frac 1 D \geq \mathrm{ess} \limsup_{n \to \infty} \frac{1 -  \A[n]^2}{\sigma[n]^2}$.
\label{thm:rddiscrete}
\end{thm}

The following result on the compression of a single random variable will be key in the proof of Theorem~\ref{thm:rddiscrete}.
\begin{thm}[conditional Gaussian distortion-rate function]
\label{thm:grd}
Fix an arbitrary $P_U$, where $U$ is a random variable defined on an abstract alphabet $\mathcal U$. Conditioned on $U = u$, where $u \in \mathcal U$, let $X$ be distributed as $\mathcal N(\mu_u, \sigma_u^2)$.  The conditional distortion-rate function lower bounded as
 \begin{align}
\mathbb D_{r}(X|U)&\triangleq
  \min_{\substack{ 
P_{Y | XU} \colon \\
I(X; Y |U) \leq r}} 
\E{ (X - Y)^2} \\
 &\geq  \exp \left( - 2 r + \E[\log \sigma_U^2]\right).
\end{align}
Equality is achieved if and only if $r \geq \max_u \left\{ \E[\log \sigma_U] - \log \sigma_u \right\}$, and it is achieved by $Y$ such that conditioned on $U= u$, the variable $X$ can be expressed as $X = Y + Z$, where $Z \sim \mathcal N(0, \E[\log \sigma_U^2] \exp \left( - 2 r \right))$. 
\end{thm}

\begin{proof}
From the classical scalar Gaussian rate-distortion function~\cite[Th. 10.3.2]{cover2012elements}, we know that the conditional on the realization $U = u$ distortion-rate function
\begin{align}
\!\!\!\!  \mathbb D_{r_u}(X|U = u)&\triangleq  \min_{\substack{ 
P_{Y | X, U = u} \colon \\
I(X; Y |U = u) \leq r_u}} \E[(X - Y)^2 | U = u]  \\
&= \sigma_u^2 \exp \left( - 2 r_u \right) \label{eq:Rdgclas}
\end{align}
is achieved by $Y$ such that conditioned on $U= u$, the variable $X$ can be expressed as $X = Y + Z$, where $Z \sim \mathcal N(0, \sigma_u^2 \exp \left( - 2 r_u \right))$. Using \eqref{eq:Rdgclas}, we express
\begin{align}
 \mathbb D_{r}(X|U) = \min_{r_u, u \in \mathcal U\colon \E[r_U] \leq r}\E [\sigma_U^2 \exp \left( - 2 r_U \right)].
\end{align}
Since the function $\exp(\cdot)$ is convex, Jensen's inequality yields 
\begin{align}
\!\!\!\! \E [ \exp \left( - 2 r_U  + \log \sigma_U^2 \right)] \geq  \exp \left( \!- 2 \E[r_U]  + \E[\log \sigma_U^2] \right)\!, \!\!\!
\end{align}
with equality if and only if $ r_u  =  r - \E[\log \sigma_U] + \log \sigma_u$.
\end{proof}

\begin{proof}[Proof of Theorem~\ref{thm:rddiscrete}]
We first show the lower bound \eqref{eq:Rdlb} and then show an estimation policy that achieves it. 
Fix arbitrary estimation $\left\{\p_{\hX^k \| X^k, U^{k-1}}\right\}_{k = 1}^\infty$ and control $\left\{\p_{U^k \| \hat X^k }\right\}_{k = 1}^\infty$ policies satisfying \eqref{eq:main_const}.

For an arbitrary $X$, denote the random variable
\begin{align}
 \!\!\!\!\! V_{k-1}[X] \triangleq \E[ (X - \E[X | \hat X^{k-1}, U^{k-1}  ])^2 |  \hat X^{k-1}, U^{k-1}  ].
\end{align}
Using \eqref{eq:discrete-dynamics}, we note the relationship
\begin{align}
  V_{k-1}[X[k]]  &= V_{k-1}[ \A[k-1] X[k-1] +  \f[k-1] \\
  &\phantom{=}+ W[k-1]] \notag\\
  &= \A[k-1]^2 V_{k-1}[X[k-1]] + \sigma[k-1]^2 \label{eq:vrec}
\end{align}
Denote the per-step information rates
\begin{equation}
r[k] \triangleq I(X^k; \hat X[k] | \hat X^{k-1}, U^{k-1}  ). \label{eq:ri}
\end{equation}
Since 
\begin{equation}
 r[k] \geq  I(X[k]; \hat X[k] | \hat X^{k-1}, U^{k-1})
 \label{eq:rklb}
\end{equation}
(with equality if and only if $X^{k-1} - (X[k], \hat X^{k-1}, U^{k-1}) - \hat X[k]$), we apply Theorem~\ref{thm:grd} to lower-bound the distortion at step $k$ as   
\begin{align}
 \E[(X[k] - \hat X[k])^2 ]  &\geq  \E\left[V_{k}[X[k]]\right] \\
 &\geq \ushort d[k] \label{eq:drlb2} \\
 &\triangleq  \exp \left( - 2 r[k] + \E[\log  \Var{ V_{k-1}[X[k]] } \right) \notag
\end{align}
Combining \eqref{eq:vrec} and \eqref{eq:drlb2}, we obtain the recursion
\begin{align}
 \ushort d[k] &\geq \label{eq:recur}  \\
 &\exp \left( - 2 r[k] + \E[\log  ( \A[k-1]^2 \ushort d[k-1] + \sigma[k-1]^2) \right) \notag
\end{align}
From
$
r[k] \geq \frac 1 2 \E\left[ \log (  \A[k-1]^2 \ushort d[k-1] + \sigma[k-1]^2) \right]-$$ \frac 1 2 \log \ushort d[k]$, 
 we deduce 
\begin{align}
& \sum_{k = 1}^{n} r[k] \geq   \label{eq:sumr} \\
& \frac {1} 2 \E \left[ \log \frac{\E[ \X[ 1]^2 ]  } { \A[n]^2 \ushort d[n] + \sigma[n]^2}   +  \sum_{k = 1}^{n} \log \left(  \A[k]^2 + \frac {\sigma[k]^2}{\ushort d[k] }\right) \right]. \notag
\end{align}
Due to the distortion constraint \eqref{eq:main_const}, for any $\epsilon > 0$ there exists a $k_0$ such that for all $k > k_0$, $\ushort d[k] < D + \epsilon$.  We break up the sum \eqref{eq:sumr} into two parts, up to $k_0 - 1$ and from $k_0$ to $n$. The first part is a constant as a function of $n$, and we replace $\ushort d[k]$ by $D + \epsilon$ in the second sum since each term is decreasing in $\ushort d[k]$. To conclude \eqref{eq:Rdlb}, we normalize by $n$, take the limsup in $n$ and the limit in $\epsilon \to 0$. 

Let $D$ be small enough to satisfy the condition for the achievability in Theorem~\ref{thm:rddiscrete}. To show achievability, we construct $P_{\hat X[k] | X^{k},  \hat X^{k -1}, U^{k-1}} = P_{\hat X[k] | X[k],  \hat X^{k -1}, U^{k-1}}$ (so that \eqref{eq:rklb} is satisfied with equality) such that the a posteriori estimation errors $Z[k] \triangleq X[k] - \hat X[k]$ are independent $\mathcal N(0, D)$ conditioned on $\hat X^{k-1}, U^{k-1}$. Due to the condition for equality in Theorem~\ref{thm:grd}, such a process will satisfy \eqref{eq:drlb2} and \eqref{eq:sumr} with equality. We create an auxiliary process
\begin{equation}
Y[k] = X[k] + V[k],   
\end{equation}
where $V[k] \sim \mathcal N(0, \rho[k]^2)$ conditioned on $Y^{k-1}, U^{k-1}$, 
and we let 
\begin{equation}
\hat X[k] =  \E\left[ X[k] | Y^{k}, U^{k-1} \right]
\label{eq:mmse}
\end{equation}
be the MMSE estimate of $X[k]$ based on the observation of $Y^{k}$ and the knowledge of the control signals $U^{k-1}$.  
That the a posteriori estimation errors $\left\{Z[k]\right\}$ are Gaussian conditioned on $\hat X^{k-1}, U^{k-1}$ is known from the Kalman filter, and the parameter $\rho[k]$ is set to achieve the variance of $Z[k]$ equal to $D$ from the corresponding Riccati recursion.
\end{proof}

\subsection{ Rate-distortion function for the continuous-time system: proof of Theorem~\ref{thm:Rd}}
We first show the lower bound \eqref{eq:br-form} and then present an estimation policy that achieves it. 
For $T > 0$ and $\delta > 0$, denote for brevity $n \triangleq \lfloor  T/ \de \rfloor $ and write
\begin{align}
\mathbb R_e(D) &\triangleq \inf \limsup_{T \rightarrow \infty}  \frac{ 1 } {T }  I (   \xdev^T \rightarrow \hX^T   \| U^{T-} ) 
\label{eq:cont_uc_lim_1} \\
&=\inf \limsup_{T \rightarrow \infty} \lim_{\de \rightarrow 0}    \frac{1}{\de n}  I ( \X^{n} \rightarrow \hX^{n} \| U^{ n-1} )  
\label{eq:cont_uc_lim_2}\\
&\geq \inf_{ \substack{\left\{D(t)\right\}_{t \in \mathbb R_+} \colon \\ \limsup_{t \to \infty} D(t) \leq D }}\limsup_{T \rightarrow \infty}  \lim_{\de \rightarrow 0}    \frac{1}{\de}   \notag\\
&\inf_{\substack{\p_{\hX^n \| X^n, U^{n-1}} \colon \\ \E{(X[k] - \hat X[k])^2} \leq D(\delta k),\\ k = 1, \ldots, n}}  \frac{ 1 } {n }  I ( \X^\T \rightarrow \hX^\T \| U^{\T-1} )  \label{eq:cont_uc_lim_3} \\
&\geq \inf_{ \substack{\left\{D(t)\right\}_{t \in \mathbb R_+} \colon \\ \limsup_{t \to \infty} D(t) \leq D } } \limsup_{T \rightarrow \infty} \lim_{\de \rightarrow 0}  \frac{1}{\de}   
 \frac {1} {2n} \notag\\
&\phantom{\geq} \E \Bigg[ \log \frac{\E[ \X[ 1]^2 ]  } { \A[n]^2 D(\delta n) + \sigma[n]^2}   \notag\\
 &\phantom{\E \Bigg[}
 + \sum_{k = 1}^{n} \log \left(  \A[k]^2 + \frac {\sigma[k]^2}{D(\delta k)}\right) \Bigg] 
\label{eq:cont_uc_lim_4} \\
&\geq \limsup_{T \rightarrow \infty} \frac{1}{T}\E \left[ \int_{0}^{T}  \left( -\A(\tau)  + \frac{  \sigma(\tau)^2 }{2 D}  \right) d\tau \right] ,   
\label{eq:cont_uc_lim_5} \\
&=  \frac{ \E[\sigma^2] }{2 D} - \E[\A]  
\label{eq:cont_uc_lim_6} 
\end{align}
where 
\begin{itemize}
 \item \eqref{eq:cont_uc_lim_1} is by Definition~\ref{def:causaldirectedinfo} of rate-distortion function;
 \item \eqref{eq:cont_uc_lim_2} is by Definition~\ref{def:rate_cost_estimation_continuous} of continuous-time directed information;
 \item \eqref{eq:cont_uc_lim_3} rewrites the $\inf$ over infinite-horizon estimation policies $\left\{\p_{\hX^t \| X^t, U^{t-}}\right\}_{t \in \reals_+}$ satisfying \eqref{eq:staterr} as 
 $$\inf_{ \substack{\left\{D(t)\right\}_{t \in \mathbb R_+} \colon \\ \limsup_{t \to \infty} D(t) \leq D }} ~\inf_{\substack{\left\{\p_{\hX^t \| X^t, U^{t-}}\right\}_{t \in \mathbb R_+}\colon \\ \E{(X(t) - \hat X(t))^2} \leq D(t)}}$$
 and interchanges the second $\inf$ and the limits;
 \item \eqref{eq:cont_uc_lim_4} applies \eqref{eq:sumr};
 \item \eqref{eq:cont_uc_lim_5} applies $e^x \geq 1 + x$ for $x \geq -1$, $\log(1 + x) \geq x - x^2$ for $x \geq -0.68$, and assumption \eqref{eq:ubound} on uniform boundedness of $\deathx(t)$  to deduce that for sufficiently small~$\delta$,
 \begin{align}
\deathx[k]^2 &\geq 1 - 2 \int_{(k - 1) \delta}^{k \delta}  \deathx(\tau) d\tau,  \label{eq:seca}\\
\sigma[k]^2 &\geq (1 -  2 \delta \A_{\max}) \int_{(k - 1) \delta}^{k \delta} \sigma(\tau)^2 d\tau,\label{eq:secb}
\end{align}
\begin{align}
&~\log \left( \A[k]^2 + \frac {\sigma[k]^2}{D[k] }\right) \geq  \label{eq:sec} \\
&~ \int_{(k - 1) \delta}^{k \delta}  \left( - 2\A(\tau) + \frac {1 -  2 \delta \A_{\max}} {D[k]} \sigma(\tau)^2 \right) d \tau \notag\\
-&~ \left( \int_{(k - 1) \delta}^{k \delta}  \left( - 2\A(\tau) + \frac {1 -  2 \delta \A_{\max}} {D[k]} \sigma(\tau)^2 \right) d \tau \right)^2, \notag
\end{align}
and applies $\left(\int_{a}^b f(x) dx\right)^2 \leq  (b-a) \int_{a}^b f(x)^2 dx$ to bound the second term in the right side of \eqref{eq:sec} by
\begin{align}
 \delta \int_{(k - 1) \delta}^{k \delta}  \left( - 2\A(\tau) + \frac {1 -  2 \delta \A_{\max}} {D[k]} \sigma(\tau)^2 \right)^2 d \tau,
\end{align}
concluding that this term can be ignored after summing with respect to $k$ and taking the limit in $\delta \to 0$ due to the uniform boundedness of $\A(t)$ \eqref{eq:ubound} and the existence of the 4th moment of $\sigma(t)$ for $t$ large enough \eqref{eq:4bound}.

\end{itemize}

To show the achievability of \eqref{eq:cont_uc_lim_4}, let $D$ be small enough so that the condition for the achievability in Theorem~\ref{thm:Rd} is satisfied. Following the same reasoning as in the achievability of \eqref{eq:sumr}, it is enough to manifest $P_{\hat X(t) | X^{t},  \hat X^{t-}, U^{t-}} = P_{\hat X(t) | X(t),  \hat X^{t-}, U^{t-}}$ such that the a posteriori estimation errors $Z(t) \triangleq X(t) - \hat X(t)$ are independent $\mathcal N(0, p_t)$ conditioned on $\hat X^{t-}, U^{t-}$, where $p_t \to D$. Consider an auxiliary continuous-time process
\begin{equation}
dY = X dt + \rho(t) dW,   
\label{eq:AWGN}
\end{equation}
where $W(t)$ is a Wiener process, and $\rho(t)$, which is a function of $Y^{t-}, U^{t-}$ \eqref{eq:U}, will be shown shortly. 
We let 
\begin{equation}
\hat X(t) =  \E\left[ X(t) | Y^{t}, U^{t-} \right]
\label{eq:mmsec}
\end{equation}
be the MMSE estimate of $X(t)$ based on the observation of $Y^{t}$ and the knowledge of the control signals $U^{t-}$.  
The estimation errors $\left\{Z(t)\right\}$ are Gaussian conditioned on $\hat X^{t-}, U^{t-}$, and the parameter $\rho(t)$ is set to achieve the variance of $Z(t)$ equal to $D$ using the Kalman filter recursion~\cite[Ch. 3]{lewis2017optimal} as follows. The MMSE estimate process is
 \begin{align}
 d\hat X &=  -\deathx \hat X dt +  \f dt + K d \tilde Y \label{eq:kalmanc},
 \end{align}
where 
\begin{align}
 d \tilde Y &\triangleq dY -  \A \hat X dt
\end{align}
is the {\it innovation} process, 
and the Kalman filter gain $K(t)$  is given by
\begin{align}
K(t) &\triangleq \frac{p(t)}{\rho(t)^2}, 
\label{eq:K(t)}
\end{align}
where variance $p(t)$ of the a priori estimation error given $Y^{t-}, U^{t-}$ and the auxiliary channel noise power $\rho(t)$ are found from the Riccati recursion for $p(t)$: 
\begin{align}
\frac{dp}{dt} &= (-2 \deathx(t)  - K(t)) p(t)  + \sigma(t)^2 \label{eq:Pt+1|tc},\\
p(0) &= \Var{X(0)}, ~p(\infty) = D,
\end{align}
where we equated the stationary variance of $p(t)$ to the desired value $D$. 
To show the achievability of \eqref{eq:cont_uc_lim_5}, we apply a Taylor series expansion to reverse the inequalities \eqref{eq:seca}, \eqref{eq:secb} and \eqref{eq:sec} up to the addition of a $O(\delta^2)$ term. That remainder term is uniformly bounded in $k$ due to the assumptions \eqref{eq:ubound} and \eqref{eq:4bound}. 
 \qed

\section{Biomolecular control: proofs}
\label{apx:biomolecular}
\begin{proof}[Proof of \eqref{eq:rdcbio}]
If the expectation of $X(t)$ converges to a unique stationary value as implied by the control objective \eqref{eq:fano}, then 
 \begin{equation}
 \label{eq:sonservation}
\hspace*{-4mm}
 \E[ \birthx ] = \E[ \deathx X ].
\end{equation}
The bound follows by plugging \eqref{eq:one_def}, \eqref{eq:little}, and \eqref{eq:sonservation} into \eqref{eq:rdcbound}.
\end{proof}

\begin{proof}[Proof of Proposition~\ref{prop:Fano}]
We consider the discretized system \eqref{eq:discrete-dynamics}, and we set the channel input as $V[k] = \alpha[k](X[k] - \bar X[k])$, where $ \bar X[k]$ is the controller's prediction of the state $X[k]$ based on $Y[1], \ldots, Y[k-1]$, and $\alpha[k]$ is a factor set to satisfy the power constraint of the channel. The controller applies the Kalman filter to generate $\hat X[k]$, the MMSE estimate of $X[k]$, where the estimation errors $Z[k] \triangleq X[k] - \hat X[k]$ are zero-mean Gaussian independent of $\hat X[1], \ldots, \hat X[k]$.
Assuming without loss of generality that the objective is to steer the system to location 0, the optimal controller applies the additive control input as
\begin{align}
\lambda[k] = - \mu[k] \hat X[k],
\end{align}
resulting in the mean-square deviation from 0
\begin{align}
\E\left[ X[k+1]^2\right] = \E\left[ \mu[k]^2  Z[k] ^2 \right] + \sigma[k]^2
\label{eq:discreterec}
\end{align}

This encoder-controller pair is optimal in the strong sense that it achieves the (discrete-time counterpart of)  the converse \eqref{eq:R-and-Y} with equality. Under the formalism of discrete time, this remarkable matching of the Gauss-Markov source to the AWGN channel has been noted in \cite{TatikondaSahaiMitter,freudenberg2010stabilization,khina2016multi,kostina2016rate}. Via a continuous-time limit argument, as in Appendix~\ref{app:thm1}, we conclude that \eqref{eq:rdcbio} is achieved as well.

Consider \eqref{eq:discreterec}. Since the estimation errors are independent of $\hat X[k]$, we have that for any $\mu[k]$ that is a function of $\hat X[1], \ldots, \hat X[k]$ only, 
\begin{align}
\E\left[ \mu[k]^2  Z[k] ^2 \right] = \E\left[ \mu[k]^2 \right] \E \left[ Z[k] ^2 \right],
\end{align}
and that the left side of \eqref{eq:discreterec} is minimized by setting $\E\left[ \mu[k]^2 \right]$ as small as possible under the constraint \eqref{eq:ubound}, which corresponds to $\mu[k] \geq a$, for some $a > 0$. This yields $\mu[k] \equiv a$, i.e., constant degradation rate. 
\end{proof}

\bibliography{yorie.bib}
\bibliographystyle{IEEEtran}

\end{document}

%% file: macro.tex
\newtheorem{thm}{\textbf{Theorem}}

\newtheorem{prop}{\textbf{Proposition}}
\newtheorem{defn}{\textbf{Definition}}

\newcommand{\dt}{h}
\newcommand{\xgoal}{x^\star}
\newcommand{\xdev}{X}

\newcommand{\birthx}{\lambda}
\newcommand{\deathx}{\mu}

\newcommand{\A}{\mu}			
\newcommand{\f}{\lambda}	

\newcommand{\reals}{\mathbb{R}}
\newcommand{\ints}{\mathbb{Z}}

\newcommand{\Pro}{\mathbb{P}}
\newcommand{\E}{\mathbb{E}}
\providecommand{\var}{\mathrm{Var}}

\newcommand{\mean}{\mathbb{E}}
\newcommand{\svar}{\mathrm{Var}}

\newcommand{\one}{\gamma_{X}}

\newcommand{\ie}{i.e.}

\newcommand{\de}{\delta}

\newcommand{\p}{P}

\newcommand{\capa}{\mathbb{C}}

\newcommand{\X}{X}

\newcommand{\hX}{\hat{X}}
\newcommand{\T}{n}

\newcommand{\ck}[1]{\textcolor{black}{#1}}

\newcommand{\Var}[1]{\mathrm{Var}\left[#1\right]}